%% file: main.tex
\documentclass[conference,10pt]{IEEEtran}

\newcommand{\add}[1]{#1}

\usepackage{soul}

\IEEEoverridecommandlockouts
\usepackage{amsmath,amssymb,amsfonts}
\usepackage{mathtools}
\usepackage{algorithmic}
\usepackage{graphicx}
\usepackage{textcomp}
\usepackage{xcolor}
\usepackage{float}
\usepackage{gensymb}

\usepackage[bookmarks=false]{hyperref}
\usepackage{cleveref}

\usepackage[style=ieee]{biblatex}

\def\BibTeX{{\rm B\kern-.05em{\sc i\kern-.025em b}\kern-.08em
    T\kern-.1667em\lower.7ex\hbox{E}\kern-.125emX}}

\begin{document}

%TODO: IEEE Transaction is not a conference title
%TODO: Gå igenom och använd konsekvent NCR eller repeater. Helst NCR?
%TODO: Ska man ha eq-nummer på allt?

\title{Impact of \add{Network-Controlled Repeaters} in Integrated Sensing and Communication Systems}

\author{\IEEEauthorblockN{Henrik Åkesson, Diana P. M. Osorio, Erik G. Larsson}
\IEEEauthorblockA{\textit{Communication Systems}, \textit{Linköping University}, Linköping, Sweden \\
Emails: \{henrik.akesson, diana.moya.osorio, erik.g.larsson\}@liu.se}
}

\maketitle
% \egl{Title: use ``network controlled repeater'' instead of ``smart repeater''? (I also like ``swarm repeaters'' if there are many}

\begin{abstract}
    Integrating sensing capabilities into existing massive MIMO communication networks has become crucial, stemming from a need for a more interconnected society. Improved coverage and performance can be obtained by incorporating new network components, such as reconfigurable intelligent surfaces or network-controlled repeaters (NCR). Integrating such components into modern networks brings a number of challenges. Thus, this paper contributes with the analysis of NCR impact in integrated sensing and communication networks. Particularly, the Cramér-Rao bound for a range estimator is derived where the interference from the repeater is taken into consideration. Additionally, a joint procedure for determining the repeater amplification factor, along with the precoders of the transmitting access point, is proposed. 
\end{abstract}


% flytta ofdm till problem formulation (kortare version)
% introduction: förklara varför repeaters är bra för mimo-kommunikation, steg 1: förklara att det inte undersökts för pverkan på ISAC, steg 2: förklara inte undersökt använda för ISAC, förklara att denna artikel den första. Nämn RIS i ISAC? men NCR inte nämnts lika mycket

\section{Introduction}

\input{sections/introduction}

\section{System Model}
\input{sections/system-model}

\section{User SINR}
\input{sections/ue-sinr}

\section{Cramér-Rao Bound}
\input{sections/crb}

\section{Results}
\input{sections/results}

\section{Conclusions}
\input{sections/conclusions}

% %\section*{References}
\printbibliography

\end{document}

%% file: sections/introduction.tex
In the evolution of multiple antenna technologies towards the next generation of wireless networks, it is expected that large arrays in massive multiple-input multiple-output (MIMO) systems as well as distributed implementations will be crucial to provide the demands on connectivity. However, to overcome blind spots, specially at the cell edge, new network components have emerged to overcome the challenges of the propagation environment and provide enhanced coverage and performance, such as reconfigurable intelligent surfaces (RIS) and
network-controlled repeaters (NCR) \cite{ayoubi_network-controlled_2023}.

\add{
While NCR has been already standardized in 3GPP Release 18 \cite{3GPP_TR}, and it is expected to continue evolving, the future standardization for RIS is still uncertain. Nonetheless, there has been a growing number of works demonstrating the effectiveness and capabilities of RIS in controlling the propagation environment~\cite{basar_reconfigurable_2024}. However, the benefits of NCR or RIS depends on the specific network needs. 
Besides extending the network, these components are attractive for being an 
energy efficient, and more environmentally friendly, alternative to conventional 
cellular networks \cite{abedini_smart_2024}, which is crucial to address current global challenges. 

However, considering urban scenarios where quick and easy deployment is essential, NCR proves to provide a flexible way to achieve the benefits of distributed MIMO, while at the same time being cheaper and less demanding to deploy \cite{willhammar_achieving_2024}. In \cite{da_silva_impact_2024}, it is even 
suggested that a very small number of NCR, with sufficient network planning, may
notably improve quality of service for users who would be in large fading environments
with respect to the access points (AP). From a power consumption perspective, NCRs can even outperform RIS by reducing the overall network energy consumption in certain scenarios \cite{abedini_smart_2024}.

On the other hand, networks are evolving to incorporate sensing capabilities, thereby realizing integrated sensing and communication (ISAC) systems, to enable a range of advanced use-cases including autonomous driving and positioning \cite{gonzalez-prelcic_integrated_2024}. It is intuitively that these new network components will defenitely play a role to support the performance of both capabilities, communication and sensing. Indeed, research on RIS to support ISAC has received a recent increasing attention~\cite{ismail_ris-assisted_2024,sheemar_full-duplex-enabled_2023}, while the role of NCR in ISAC systems has been neglected. 

Acknowledging the significant potential of NCR in enhancing ISAC networks, we aim to address the existing research gap in this area. Specifically, we identify two critical questions: what is the impact of deploying NCR to assist communications over the sensing task, and how can NCRs be leveraged to improve sensing performance while maintaining the required communication quality? This paper will focus on the first question by tackling the fundamental task of analyzing the potential impact on sensing by the deployment of an NCR to extend the communication coverage. 
}

\subsection{Problem Formulation}
Herein, it is considered an amplify-and-forward repeater that operates in full-duplex mode and is capable of canceling the self interference. However,  non-cancelable interference will be
present in what is inadvertently amplified and transmitted back to the AP. Although the scenario considers only one repeater, it is worth mentioning that in networks with multiple repeaters, positive feedback between repeaters can disrupt stability and lead to system failure~\cite{larsson_stability_2024}.
% \egl{I think
% actually if the gain passes beyond the stability limit then nothing
% will work... at all -- repeaters will be saturated and not transmit
% anything useful.}

\add{Considering its widespread use in modern communication networks, orthogonal frequency-division multiplexing (OFDM) is utilized as ISAC waveform in this work. Although
it is not a waveform designed taking radar into consideration, multiple previous works highlight its potential usefulness \cite{sakhnini_cramer-rao_2021,sit_mimo_2014}. Particularly, it has been shown that the spectral structure of the OFDM signal subsidizes improved detection capabilities in a system with multiple moving targets, by exploiting the inter-carrier interference caused by the Doppler shifts \cite{keskin_mimo-ofdm_2021}.}
%sit_mimo_2014,
To reduce the interference from the repeater, on the sensing endeavors of the AP, we formulate the minimization of the Cramér-Rao bound ($\mathrm{CRB}$) \cite{Kay97} of a range estimate to a radar target \cite{wang_cramer-rao_2009} by optimizing the $M$-antenna-AP precoder and the repeater gain. \add{Moreover, it is considered a monostatic radar system, i.e., the transmitter and the receiver are co-located in the AP. Furthermore, the repeater and the AP are assumed to be connected through a centralized control unit, capable of adjusting the gain of the NCR.}

% \add{We limit the scope of this paper to this relatively simple estimate, in order to highlight
% the main objective of examining the NCR impact on the system, whereas e.g., angle and velocity estimates
% might be more intereseting from a MIMO-OFDM ISAC perspective \cite{wang_cramer-rao_2023, sakhnini_cramer-rao_2021}.

\subsection{Contributions}
The contributions of this paper include a closed form expression, for
the specific circumstances this paper takes into consideration, of the
$\mathrm{CRB}$ of the range estimate, within a system where the sensing is
impaired by the presence of a NCR. Moreover, we also propose a joint
optimization method for determining the amplification gain of the
repeater, while also calculating the precoder for the AP.

%% file: sections/system-model.tex
% !TEX root = ../main.tex
\begin{figure}[bt]
    \centering
    \includegraphics[width=1\linewidth]{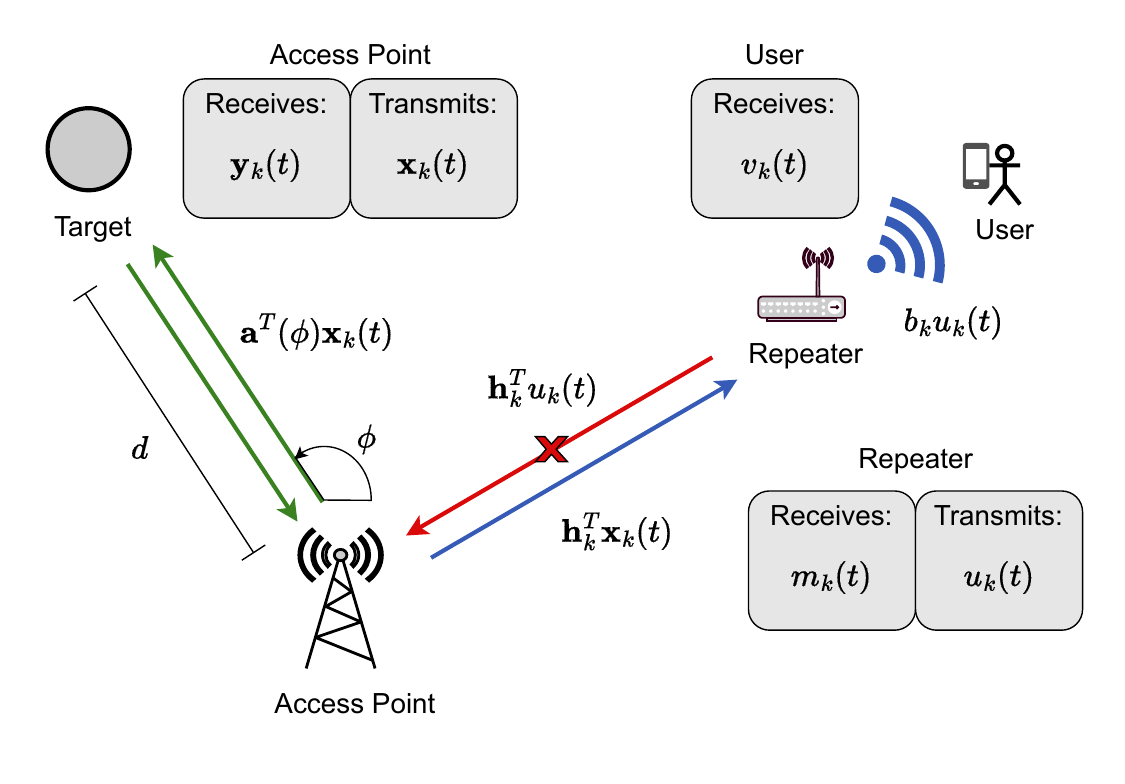}
    \caption{System model for each sub-carrier $k=0,1,2,\dots,N_s-1$}
    \label{fig:system-model}
\end{figure}

The system depicted in~\Cref{fig:system-model} illustrates the downlink communication between one $M$-antenna ISAC-AP and one single antenna user assisted by a single antenna NCR, as no direct link between the AP and user exists. Additionally, the AP is sensing one stationary target whose distance from the AP is unknown. All components of additive white Gaussian noise (AWGN) follow a $\mathcal{CN}(0, \sigma_{e}^2)$ distribution, where $\mathcal{CN}$ denotes a circularly-symmetric complex Gaussian distribution with zero mean and $\sigma_{e}^2$ variance. In the case of noise vectors, all vector entries follow the aforementioned distribution. 

\subsection{Signal models}
It is considered an OFDM signal with symbol duration of $T_{tot}$ s.t. $0 \leq t < T_{tot}$, where $T_{tot} = T_{symb} + T_{CP}$, which is the time duration of a data symbol plus the time duration of the cyclic prefix. %For specifics regarding the OFDM radar aspects, \cite{braun_ofdm_2014} covers the fundamentals.

\subsubsection{Transmitted signal from the AP}
The $M\times 1$ vector transmitted from the AP at time $t$ is 
\begin{equation}\label{eq:base_tx_signal}
    \mathbf{x}(t) = \mathbf{w}s(t),
\end{equation}
\noindent where $\mathbf{w}$ is the $M \times 1$ precoding vector, and $s(t)$ is the OFDM symbols
\begin{equation}\label{eq:ofdm_structure}
    s(t) = \sum\limits_{k=0}^{N_s - 1}c_ke^{j2\pi f_kt}.
\end{equation}
\noindent In \eqref{eq:ofdm_structure}, $N_s$ is the number of sub-carriers, and $f_k$ is the sub-carrier frequency. The $c_k$ are the complex symbols to be transmitted, which adhere to the constraint: $|c_k|=1$, i.e., the symbols have unit energy.

Additionally, the transmitted signal for a specific sub-carrier $k$ can be written as
\begin{equation}
     \mathbf{x}_k(t) = \mathbf{w}c_ke^{j2\pi f_kt},
\end{equation}
\noindent where again, $c_k$ is a complex scalar representing one data symbol, for sub-carrier $k$.

\subsubsection{Received and transmitted signal at the repeater}

The received signal at the repeater, for sub-carrier $k$, is modeled as 
\begin{equation}\label{eq:repeater_received}
    m_{k}(t) = \mathbf{h}_{k}^T\mathbf{x}_k(t) + n_{rk}(t),
\end{equation}
\noindent where $\mathbf{h}_{k}$ is the $M \times 1$ complex channel coefficient vector between the AP and the repeater, and $n_{rk}(t)$ is AWGN.
% \egl{Is the Rayleigh assumption really needed here for this equation to hold?}

Then, following an amplify-and-forward procedure with amplification gain $\alpha$, the repeater transmits
\begin{equation}
    u_{k}(t) = \alpha m_k(t).
\end{equation}
% \egl{Consider a different symbol than $x$ for this repeater-transmitted signal. The notation $x_k$ seems ambiguous, is it the
% $k$th component of $\textbf{x}$ or is it the repeater signal}

\subsubsection{Received signal at the AP}
Considering a monostatic radar system consisting of a single ISAC AP, 
% \egl{say that earlier, it was maybe obvious but perhaps say it explicitly upfront} 
we can define $\mathbf{a}(\phi) $$=$$ \left[1 \quad e^{j\pi\cos{\phi}}\quad e^{j2\pi\cos{\phi}}\quad \dots \quad e^{j(M-1)\pi\cos{\phi}}\right]^T$ to be the antenna array response vector for an angle of arrival/departure $\phi$, towards the radar target, which allows us to write the received signal at the AP as  
\begin{equation}\label{eq:received_mimo_radar}
    \mathbf{y}_k(t) = \sigma\add{\beta(d)}\mathbf{a}(\phi)\mathbf{a}(\phi)^T\mathbf{x}_k(t-\tau(d)) + \mathbf{h}_{k} u_k(t) + \mathbf{n}_k(t).
\end{equation}
\noindent In \eqref{eq:received_mimo_radar}, all vectors are $M\times 1$ and:
\begin{itemize}
    \item $\sigma$ is the RCS which is distributed according to $\mathcal{CN}(0, \sigma_{\text{RCS}}^2)$
    \item $\beta(d)$ is the large scale fading of the radar echo which depends on the distance to the radar target. Since we consider line of sight to the radar target, the path loss is modeled as $\beta(d) = \dfrac{1}{d^2}$
    \item $\tau(d)$ is the delay introduced from the round-trip path from the AP to the radar target, and back
    \item $\mathbf{h}_k$ is the complex channel coefficient vector between the AP and the repeater. Note that due to the reciprocity of the channel, this vector is equals the one in \eqref{eq:repeater_received}
    % \egl{So this means we get the repeater roundtrip channel $h h^H$? Should it be $h h^T$?}
    \item $\mathbf{n}_k$(t) is an AWGN noise vector
\end{itemize}

\subsubsection{Received signal at user}
The received signal at the user is
\begin{equation}\label{eq:received_user}
    v_k(t) = b_ku_k(t) + n_{uk}(t),
\end{equation}
\noindent where $b_{k}$ is the complex channel coefficient between the repeater and the user, and $n_{uk}(t)$ is AWGN.
% \egl{I feel there is a notation ambiguity here, $h_k$ could be mistaken for the $k$th component of $\textbf{h}$.
% Consider using another symbol. Maybe all of this would become clear with a good figure that introduces
% notation}

In the following, we will consider sampled signals, since this will allow simple forms for the expressions in the later sections. In~\eqref{eq:ofdm_structure}, if we set $f_k =$$ k\Delta_f$ for $k = 0, 1,\dots,N_s-1$, and a sub-carrier spacing of $\Delta_f$, then \eqref{eq:ofdm_structure} can be written as
\begin{equation}\label{eq:ofdm_structure2}
    s(t) = \sum\limits_{k=0}^{N_s - 1}c_ke^{j2\pi k\Delta_ft},
\end{equation}
\noindent and, after sampling with frequency $f_s=\Delta_f N_s$, we have that
\begin{equation}\label{eq:ofdm_structure2_sampled}
    s\left(\dfrac{n}{f_s}\right) = s[n] = \sum\limits_{k=0}^{N_s - 1}c_ke^{j2\pi k\frac{n}{N_s}}.
\end{equation}
\noindent Then, we can express $\mathbf{x}[n] $$=$$ \mathbf{w} s[n],$ and $\mathbf{x}_k[n] $$= $$\mathbf{w}c_ke^{j2\pi k \frac{n}{N_s}}.$ Thus, by applying the same reasoning, we rewrite \eqref{eq:received_mimo_radar} as 
\begin{multline}\label{eq:received_mimo_radar_sampled_sc}
    \hspace{-10pt} \mathbf{y}_k[n] = \sigma\add{\beta(d)}\mathbf{a}(\phi)\mathbf{a}(\phi)^T\mathbf{x}_k\left(\frac{n}{f_s}-\tau(d)\right) + \mathbf{h}_{k} u_k[n] + \mathbf{n}_k[n] \\
    \add{= \sigma\beta(d)\mathbf{a}(\phi)\mathbf{a}(\phi)^T\mathbf{x}_k\left(\frac{n}{f_s}-\tau(d)\right)} \\ +
    \add{\alpha\mathbf{h}_{k}\mathbf{h}_{k}^T\mathbf{x}_k[n]+ \alpha \mathbf{h}_{k} n_{rk}[n] + \mathbf{n}_k[n].}
\end{multline}

% \egl{Consider if you want to write out a ``complete'' signal model, insert $x_{k}$ and $y_{rk}$ so it becomes explicit that
% we have the rank-one interference from the repeater}
\noindent Since $0 \leq t < T_{tot}$, this sampling procedure results in $N =$$ \lfloor T_{tot}f_s \rfloor$ complex samples. 
% \egl{I made the occasional language correction directly in the file, hope that's okay}

%% file: sections/ue-sinr.tex
% !TEX root = ../main.tex
Using the received signal model for the single antenna user, we can derive an expression for the signal-to-interference-plus-noise ratio (SINR), $\gamma$, by expanding \eqref{eq:received_user}. Simultaneously, we also sample the received signal at the user resulting in
\begin{align}
    v_k[n] &= b_ku_k[n] + n_{uk}[n] \\ 
    &= b_k\alpha\mathbf{h}_k^T\mathbf{x}_k[n] + b_k\alpha n_{rk}[n] + n_{uk}[n] \\
    &= b_k\alpha\mathbf{h}_k^T \mathbf{w}c_ke^{j2\pi k \frac{n}{N_s}} + b_k\alpha n_{rk}[n] + n_{uk}[n].
\end{align}

\noindent From this, the useful signal is
\begin{equation}\label{eq:sinr_effective}
    r_k[n] = b_k\alpha\mathbf{h}_k^T \mathbf{w}c_ke^{j2\pi k \frac{n}{N_s}},
\end{equation}

\noindent and the undesired terms are
\begin{equation}\label{eq:sinr_errors_terms}
    e_k[n] = b_k\alpha n_{rk}[n] + n_{uk}[n].
\end{equation}

Furthermore, we define the SINR as
\begin{equation}\label{eq:sinr_def}
    \gamma = \dfrac{|r_k|^2}{\mathbb{E}\{|e_k|^2\}},
\end{equation} 
\noindent %Noticing the fact that $n_{rk}$, and $n_{uk}$ are independent of each other, as well as between different $n$, to disregard the time index. 
thus, the unknown channel between the user and the AP, $b_k\mathbf{h}_k \coloneqq \mathbf{g}_{k}$ is estimable through uplink pilots from the user. Herein, we assume a perfect estimation is feasible, \add{considering an approximately stationary user, and thus a relatively long coherence time.}
% \st{since it is outside the scope of this paper to examine the communication aspects of the proposed system.} \egl{Consider instead saying that coherence time is long enough
% (mobility small enough) that one can estimate "perfectly"?} 
Therefore, the received useful signal power is expressed by
\begin{equation}
    |r_k|^2 = \alpha^2|\mathbf{g}_{k}^T\mathbf{w}|^2,
\end{equation}
\noindent for a specific sub-carrier $k$, and where $|c_k|=1$ is used. We also determine the denominator in \Cref{eq:sinr_def} as
\begin{equation}
    \mathbb{E}\{|e_k|^2\} = \alpha^2 \sigma_b^2 \sigma_e^2 + \sigma_e^2,
\end{equation}
\noindent where $\sigma_b^2$, and $\sigma_e^2$ are the variances of $b_k$ and the AWGN, respectively.

Subsequently, the final expression for the SINR at the user is 
\begin{equation}
    \gamma = \dfrac{\sum\limits_{k=0}^{N_s-1}\alpha^2 |\mathbf{g}_{k}^T\mathbf{w}|^2}{N_s\alpha^2 \sigma_b^2 \sigma_e^2 + N_s\sigma_e^2},
\end{equation}
\noindent and is summed over all $k$, since there is no correlation between different sub-carriers. Although we assume perfect estimation of the total channel between user and AP, the individual channels $b_k$ and $\mathbf{h}_k$ are still unknown and random.

%% file: sections/crb.tex
\add{The $\mathrm{CRB}$ of the range, $d$, to the target is denoted $\mathrm{CRB}_d$, 
and depends on the $M\times 1$ complex precoding vector $\mathbf{w}$, and the amplification gain of the NCR, $\alpha$, which is a positive real number.}
Furthermore, $\gamma_u$ is the minimum user SINR constraint, respectively. Thus, the problem can be expressed as
\begin{subequations}\label{eq:optimization_problem}
    \begin{alignat}{5}
        \min_{\mathbf{w}, \alpha} & \quad && \mathrm{CRB}_d\label{eq:sensing_centric_objective_fun}\\
        \text{s.t.} & && \gamma && \ge \gamma_u\label{eq:ue_sinr_constraint} \\
        & && ||\mathbf{w}||^2 && \le P_{max}
    \end{alignat}
\end{subequations}
\noindent where $P_{max}$ is the maximum transmit power constraint on the AP. In~\eqref{eq:received_mimo_radar_sampled_sc}, assuming that the angle towards the target, $\phi$, is known, there are three unknown parameters: $d$, $\sigma$, and $\mathbf{h}_k$. Importantly, $\beta(d)=\dfrac{1}{d^2}$ and $\tau(d)=2\dfrac{d}{c}$, where $c$ is the speed of light, contain no randomness.

The $\mathrm{CRB}_{\boldsymbol{\theta}}$, of some parameter vector $\boldsymbol{\theta}$, is determined by the inverse of the Fisher information matrix, $\mathcal{I}^{-1}(\boldsymbol{\theta})$, whose $i$th diagonal element corresponds to the $\mathrm{CRB}$ of the $i$th element of $\boldsymbol{\theta}$. 

Herein, we consider a downlink pilot transmission phase conducted by the AP, where the round-trip channel to the repeater, and back, is estimated. This estimation, denoted $\mathbf{\widehat{H}}_k$, results in an estimation error $\mathbf{\widetilde{H}}_k$ per sub-carrier $k$, consisting of i.i.d. $\mathcal{CN}(0, \sigma_{\mathbf{H}}^2)$ entries, where $\sigma_{\mathbf{H}}^2$ is the channel estimation error variance. Accordingly, by replacing $\mathbf{h}_k\mathbf{h}_k^T$ in \eqref{eq:received_mimo_radar_sampled_sc}, the received signal at the AP is
\begin{equation}\label{eq:received_per_sc}
\begin{split}
    \mathbf{y}_k[n] = & \underbrace{\sigma\mathbf{a}(\phi)\mathbf{a}(\phi)^Te^{j2\pi k\frac{n}{N_s}} \mathbf{w}c_k\beta(d)e^{j2\pi k\Delta_f\tau(d)}}_{\text{Useful signal}} \\
    & + \underbrace{\alpha\mathbf{\widehat{H}}_k \mathbf{w}c_ke^{j2\pi k\frac{n}{N_s}}}_{\text{Known and cancelable}} + \underbrace{\alpha\mathbf{\widetilde{H}}_k \mathbf{w}c_k e^{j2\pi k\frac{n}{N_s}}}_{\text{Unknown interference}} \\
    &+ \underbrace{\alpha \mathbf{h}_{k} n_{rk}[n]}_{\text{Propagated noise, } \alpha\mathbf{z}_k[n]} + \underbrace{\mathbf{n}_k[n]}_{\text{Noise}}. \\
\end{split}
\end{equation}
\noindent The term $\alpha\mathbf{h}_kn_{rk}[n]$ proves cumbersome to rigorously include in this analysis, therefore we will model this proportionally small attenuated, noise as $\alpha\mathbf{z}_k[n]$ where $\mathbf{z}_k[n] $$\sim$$ \mathcal{CN}(\mathbf{0}, \sigma_\mathbf{z}^2\mathbf{I}_M)$. Then, the precoded symbols are defined as $\mathbf{d}_k=\mathbf{w}c_k$. To facilitate the forthcoming derivations, we define the vectorized channel estimation error $\boldsymbol{\mathfrak{h}}_k= \mathrm{vec}\{\mathbf{\widetilde{H}}_k\}\sim\mathcal{CN}(\mathbf{0}, \sigma_{\mathbf{H}}^2\mathbf{I}_{M^2})$, and $\mathbf{D}_k = \mathbf{I}_M \otimes \mathbf{d}_k$, where $\otimes$ denotes the Kronecker product. The unknown interference is thus simplified to $\alpha e^{j2\pi k\frac{n}{N_s}}\mathbf{D}_k^T\boldsymbol{\mathfrak{h}}_k$. Finally, using these definitions, and removing the known interference, we end up with
\begin{multline}\label{eq:received_simplest}
    \mathbf{y}_k[n] = \sigma\mathbf{a}(\phi)\mathbf{a}(\phi)^Te^{j2\pi k\frac{n}{N_s}}\mathbf{d}_k\beta(d)e^{j2\pi k\Delta_f\tau(d)} \\
    + \alpha e^{j2\pi k\frac{n}{N_s}}\mathbf{D}_k^T\boldsymbol{\mathfrak{h}}_k + \alpha\mathbf{z}_k[n] + \mathbf{n}_k[n].
\end{multline}

Hence, we define, for each time sample $n$ and sub-carrier $k$,
\begin{multline}
    \boldsymbol{\mu}_{nk}(\boldsymbol{\xi}) = \sigma\mathbf{a}(\phi)\mathbf{a}(\phi)^Te^{j2\pi k\frac{n}{N_s}}\mathbf{d}_k\beta(d)e^{j2\pi k\Delta_f\tau(d)} \\
    = (\sigma^{\Re} + j\sigma^{\Im})\mathbf{a}(\phi)\mathbf{a}(\phi)^Te^{j2\pi k\frac{n}{N_s}}\mathbf{d}_k\beta(d)e^{j2\pi k\Delta_f\tau(d)},
\end{multline}

\noindent where $\sigma^{\Re}$ and $\sigma^{\Im}$ are the real and imaginary component of $\sigma$, respectively, and $\boldsymbol{\xi} = \left[d, \sigma^{\Re}, \sigma^{\Im}\right]$ is the parameter vector.

Similarly, the covariance of \eqref{eq:received_simplest} is
\begin{align}
    \boldsymbol{\Sigma} &= \mathrm{cov}\left\{\alpha e^{j2\pi k\frac{n}{N_s}}\mathbf{D}_k^T\boldsymbol{\mathfrak{h}}_k + \alpha\mathbf{z}_k[n] + \mathbf{n}_k[n]\right\} \\
    &= \alpha^2\sigma_{\mathbf{H}}^2\underbrace{\mathbf{D}_k^H\mathbf{D}_k}_{=(\mathbf{w}^H\mathbf{w})\mathbf{I}_M} + \alpha^2\sigma_\mathbf{z}^2\mathbf{I}_M + \sigma_e^2 \mathbf{I}_M,
\end{align}
\noindent and the subscripts $n$ and $k$ have been excluded from $\boldsymbol{\Sigma}$, due to the lack of dependence on the sub-carrier or time index.

Since the signal $\mathbf{y}_k[n]$ is complex Gaussian with distribution $\mathcal{CN}(\boldsymbol{\mu}_{nk}(\boldsymbol{\xi}),\boldsymbol{\Sigma})$, the Fisher information matrix  $\mathcal{I}(\boldsymbol{\xi})$ is given by Slepian-Bangs formula \cite{Kay97}
\begin{equation}\label{eq:slepain_definition}
    \left[\mathcal{I}(\boldsymbol{\xi})\right]_{ij} = \sum\limits_{n=0}^{N-1} \sum\limits_{k=0}^{N_s-1} 2 \Re\left\{ \dfrac{\partial \boldsymbol{\mu}_{nk}^H(\boldsymbol{\xi})}{\partial \xi_{i}} \boldsymbol{\Sigma}^{-1} \dfrac{\partial \boldsymbol{\mu}_{nk}(\boldsymbol{\xi})}{\partial \xi_{j}} \right\},
\end{equation}

\noindent where $\Re\{\cdot\}$ extracts the real part of the expression. Note that the summations over $n$ and $k$ follows from different time instances and sub-carriers being independent.

Finally, by using \eqref{eq:slepain_definition} for $i,j = 1,2,3$, we can express the Fisher information matrix as
\begin{equation}
\mathcal{I}(\boldsymbol{\xi}) = \dfrac{\psi(\mathbf{w}, \alpha)}{d^4}
    \begin{bmatrix}
        |\sigma|^2C & S_{\Re} & S_{\Im} \\
        S_{\Re} & 1 & 0 \\
        S_{\Im} & 0 & 1
    \end{bmatrix},
\end{equation}
\noindent where
\begin{equation}
    \psi(\mathbf{w}, \alpha) = \dfrac{2MN\mathbf{w}^H \mathbf{a}(\phi)\mathbf{a}^H(\phi)\mathbf{w}}{\alpha^2\sigma_{\mathbf{H}}^2\mathbf{w}^H\mathbf{w} + \alpha^2\sigma_\mathbf{z}^2 + \sigma_e^2},
\end{equation}
\begin{equation}
    C = \left(\dfrac{4N_s}{d^2} + \dfrac{16\pi^2 \Delta_f^2 N_s(N_s-1)(2N_s - 1)}{6c^2}\right),
\end{equation}
\begin{equation}
    S_{\Re} = \left(-\dfrac{2\sigma^{\Re}N_s}{d} - \dfrac{\sigma^{\Im}4\pi \Delta_f N_s(N_s-1)}{2c}\right),
\end{equation}
\begin{equation}
    S_{\Im} = \left(-\dfrac{2\sigma^{\Im}N_s}{d} + \dfrac{\sigma^{\Re}4\pi \Delta_f N_s(N_s-1)}{2c}\right).
\end{equation}
\noindent Importantly, $S_{\Re}$ and $S_{\Im}$ does not imply the real and imaginary component of $S$, but rather the co-dependence of a distance estimate, with the real and imaginary part of $\sigma$, respectively.

Thereafter, we utilize Cramer's Rule for matrix inversions to only calculate the necessary inverse element corresponding to $d$
\begin{equation}
    \left[\mathcal{I}^{-1}(\boldsymbol{\xi})\right]_{11} = \dfrac{d^{4}}{\psi(\mathbf{w}, \alpha) \left( |\sigma|^2 C - S_{\Re}^2 - S_{\Im}^2\right)},
\end{equation}

\noindent to finally express the optimization problem, in \eqref{eq:optimization_problem}, as
\begin{subequations}\label{eq:optimization_problem_final_final}
    \begin{alignat}{5}
        \min_{\mathbf{w}, \alpha} & \quad && \dfrac{d^{4}}{\psi(\mathbf{w}, \alpha) \left( |\sigma|^2 C - S_{\Re}^2 - S_{\Im}^2\right)}\label{eq:objfun_final_final}\\
        \text{s.t.} & && \dfrac{\sum\limits_{k=0}^{N_s-1}\alpha^2 |\mathbf{g}_{k}^H\mathbf{w}|^2}{N_s\alpha^2 \sigma_b^2 \sigma_e^2 + N_s\sigma_e^2} \geq \gamma_u\label{eq:ue_sinr_constraint_final_final} \\
        & && ||\mathbf{w}||^2 \leq P_{max}.
    \end{alignat}
\end{subequations}

\noindent To solve this problem, we resort to gradient based optimization, not guaranteeing globally optimal solutions.

%% file: sections/results.tex
To evaluate the proposed joint precoder and NCR amplification gain design, 2000 trials were executed to gain numerical insights in how different parameters affected the $\mathrm{CRB}$. Channel coefficients $\mathbf{g}_k$ were modeled as Rayleigh fading realizations, $\mathcal{CN}(0, \sigma_\mathbf{g}^2)$.
\begin{table}[bt]
\caption{Simulation parameters.}
    \centering
    \begin{tabular}{| c | c || c | c |}
        \hline
        \textbf{Parameter} & \textbf{Value} & \textbf{Parameter} & \textbf{Value} \\\hline
        &&&\\[-0.9em]
        $\sigma_e^2$ & $-94$ dBm & $\sigma_\mathbf{H}^2$ & $-20$ dB\\\hline
        &&&\\[-0.9em]
        $\sigma_\mathbf{z}^2$ & $-198$ dBm & $\sigma_b^2$ & $-80$ dB\\\hline
        &&&\\[-0.9em]
        $\sigma_\mathbf{g}^2$ & $-184$ dB & $\phi$ & $30\degree$\\\hline
        &&&\\[-0.9em]
        $M$ & $64$ & $N$ & $128$\\\hline
        &&&\\[-0.9em]
        $N_s$ & $128$ & $\Delta_f$ & $120$ kHz\\\hline
    \end{tabular}
    \label{tab:parameter-table}
\end{table}
 All other parameters are specified in \Cref{tab:parameter-table}. Initial values for the optimization parameters were $\alpha=1$, and a precoder with zero phase-shift and equal power for all antenna elements, i.e., $\mathbf{w}=\sqrt{\dfrac{P_{max}}{M}}\left[1,...,1\right]\in\mathbb{C}^{M\times1}$. For comparisons, the same optimization problem was solved using a fixed gain $\alpha=18.5\text{ dB}$, since this amplification gain can highly likely ensure precoders that meet the constraints for all the different parameter values used in the experiments.
\begin{figure}[bt]
    \centering
    \includegraphics[width=0.8\linewidth]{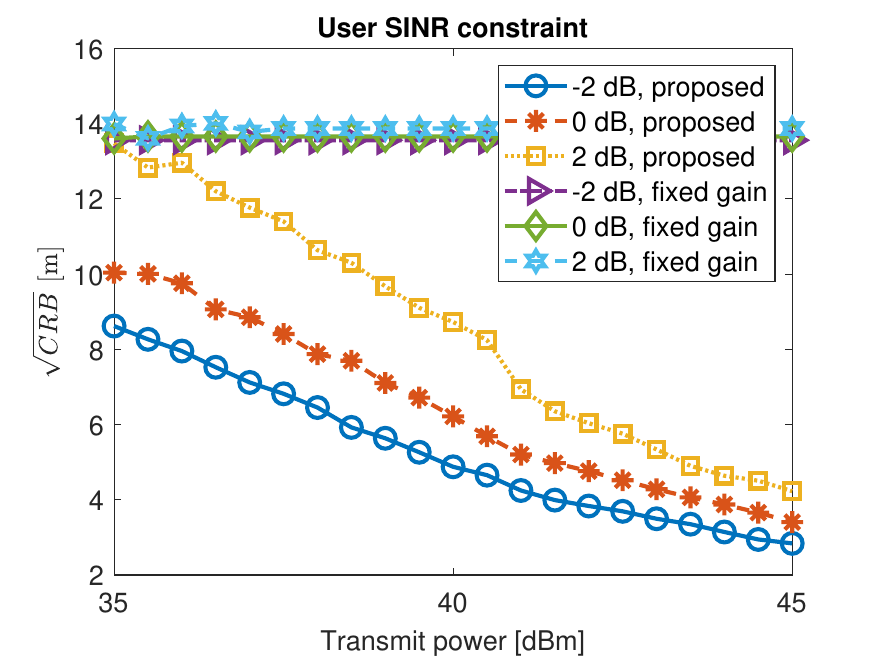}
    \caption{Effect of transmit power and user SINR constraint on the $\mathrm{CRB}$}
    \label{fig:uesinr_benchmark}
\end{figure}

\Cref{fig:uesinr_benchmark} illustrates the square root of the $\mathrm{CRB}$ versus the transmit power for different user SINR constraints. Note that, as expected, an increase in power always results in an improvement in the $\mathrm{CRB}$. Moreover, the fixed gain case presents no improvements, neither from increased power usage, nor from relaxing the use SINR constraint. This can be explained for the fact that as the power increases, the fixed gain interference from the repeater also increases as seen in \eqref{eq:objfun_final_final}. Additionally, the change in the SINR constraint implicates that less power can be required to satisfy \eqref{eq:ue_sinr_constraint_final_final}, thus behave like an increase in power for the objective function, which as before have no effect on the $\mathrm{CRB}$, for the case of a fixed $\alpha$.
\begin{figure}[bt]
    \centering
    \includegraphics[width=0.8\linewidth]{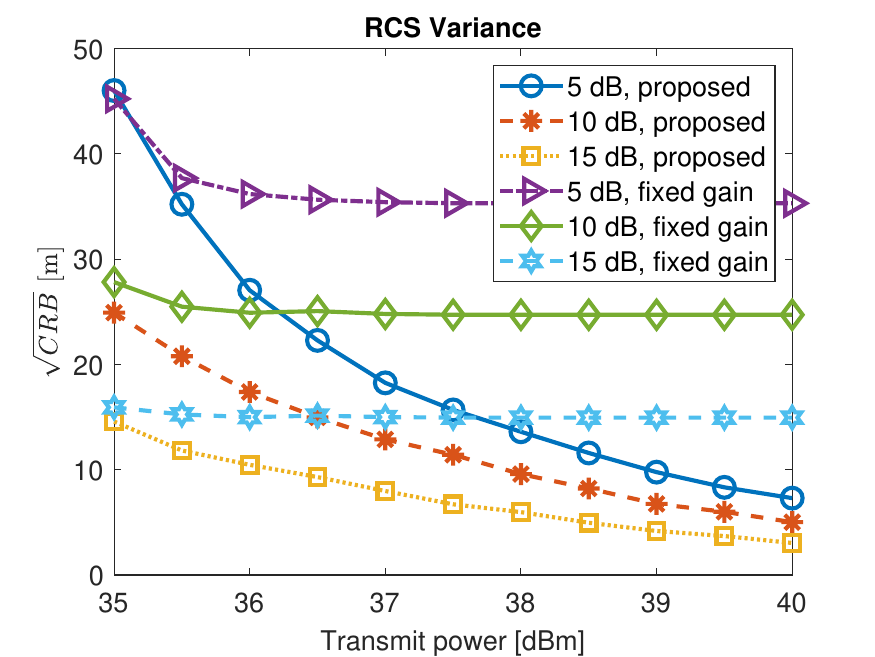}
    \caption{Effect of transmit power and RCS variance on the $\mathrm{CRB}$}
    \label{fig:rcs_benchmark}
\end{figure}

\Cref{fig:rcs_benchmark} illustrates the square root of the $\mathrm{CRB}$ versus the transmit power for three different RCS variances, for the proposed method and for the fixed amplification gain, respectively. Note that the proposed joint amplification gain and precoder optimization effectively reduces the $\mathrm{CRB}$ when the transmit power is raised. Comparing to~\Cref{fig:uesinr_benchmark}, it is observed that more transmit power do not affect the static gain case, once the RCS variance represents how visible the target is to radar sensing attempts. %, has an unsurprising effect on the $\mathrm{CRB}$.

The actual distance to the target, used in the figures presented above, was $400$ meters, and \Cref{fig:rcs_benchmark} used a user SINR constraint of $2$ dB, whilst \Cref{fig:uesinr_benchmark} used a RCS variance of $10$ dB. This results are obtained once an adjustable amplification gain will provide more space for fine tuning the system to reduce the interference from the repeater. Thus, the network is able to balance the beamforming for the communication user, with the gain of the repeater to prevent deterioration of sensing performance. An interesting future work would be to examine the opposite case, where the system favors the user in a communication-centric environment. Additionally, some areas of further analysis required is the simplified attenuated noise term in \eqref{eq:received_per_sc}, and including the estimate of the unknown channel into the $\mathrm{CRB}$ derivations.

%% file: sections/conclusions.tex
In summary, this work analyzes the effect of a NCR being used for communication coverage extension, in a MIMO-OFDM ISAC setting. Particularly, the Cramér-Rao bound was determined for a distance estimator, and an optimization problem was formulated where the amplification gain of the NCR, and the precoder vector for the AP, were the optimization parameters. There were constraints on the transmit power and minimum user SINR. Numerical results showed that jointly choosing the amplification and precoder, can reduce the minimum variance achievable by unbiased estimators further than only calculating the precoders with a fixed NCR gain.